# Direct Covalent Chemical Functionalization of Unmodified Two-Dimensional Molybdenum Disulfide

Ximo S. Chu[1], Ahmed Yousaf[2], Duo O. Li[1], Anli A. Tang[2], Abhishek Debnath[2], Duo Ma[2], Alexander A. Green[2]*, Elton J. G. Santos[3]* and Qing Hua Wang[1]*

[1]Materials Science and Engineering, School for Engineering of Matter, Transport and Energy, Arizona State University, Tempe, AZ, USA, 85287

[2]Biodesign Center for Molecular Design and Biomimetics, The Biodesign Institute, and the School of Molecular Sciences, Arizona State University, Tempe, AZ, USA, 85287

[3]School of Mathematics and Physics, Queen's University Belfast, BT7 1NN, UK

*Corresponding authors:
qhwang@asu.edu
e.santos@qub.ac.uk
alexgreen@asu.edu

**Abstract**

Two-dimensional semiconducting transition metal dichalcogenides (TMDCs) like molybdenum disulfide ($MoS_2$) are generating significant excitement due to their unique electronic, chemical, and optical properties. Covalent chemical functionalization represents a critical tool for tuning the properties of TMDCs for use in many applications. However, the chemical inertness of semiconducting TMDCs has thus far hindered the robust chemical functionalization of these materials. Previous reports have required harsh chemical treatments or converting TMDCs into metallic phases prior to covalent attachment. Here, we demonstrate the direct covalent functionalization of the basal planes of unmodified semiconducting $MoS_2$ using aryl diazonium salts without any pretreatments. Our approach preserves the semiconducting properties of $MoS_2$, results in covalent C-S bonds, is applicable to $MoS_2$ derived from a range of different synthesis methods, and enables a range of different functional groups to be tethered directly to the $MoS_2$ surface. Using density functional theory calculations including van der Waals interactions and atomic-scale scanning probe microscopy studies, we demonstrate a novel reaction mechanism in which cooperative interactions enable the functionalization to propagate along the $MoS_2$ basal plane. The flexibility of this covalent chemistry employing the diverse aryl diazonium salt family is further exploited to tether active proteins to $MoS_2$, suggesting future biological applications and demonstrating its use as a versatile and powerful chemical platform for enhancing the utility of semiconducting TMDCs.

**Introduction**

Two-dimensional semiconducting transition metal dichalcogenides (TMDCs) have elicited considerable research interest in the past few years due to their remarkable properties such as layer number dependent band gaps, photoluminescence, electroluminescence, valley polarization, and catalytic activity.[1-10] Semiconducting TMDCs are thus promising materials for electronics and optoelectronics applications.[2,3,7,11-13] For other low-dimensional materials such as carbon nanotubes and graphene, chemical functionalization enables crucial modifications of their physical, electronic, optical, and chemical properties,[14-22] and is essential for engineering how they interact with their external environment for a wide range of applications including transistors,[2,7,8,12,23] flexible electronics,[24] gas sensors,[25] and biosensors.[26-28] The chemical functionalization of TMDCs is thus expected to be similarly important for enhancing their performance in electronics and sensing applications.

Semiconducting TMDCs, however, lack dangling bonds on their basal planes,[29] which makes them substantially less reactive than the low-dimensional carbon allotropes. As a result, previous methods of covalent functionalization of semiconducting TMDCs have resorted to either harsh treatments that abolish many of the desirable properties of the TMDCs or aggressive chemicals to promote covalent bond formation.

For instance, Voiry et al.[30] and Knirsch et al.[31] have reported covalent functionalization of $MoS_2$ using organoiodides and aryl diazonium salts, respectively, only after conversion from the semiconducting 2H phase to the electron-rich metallic 1T phase. Formation of the metallic $MoS_2$ phase in these earlier works requires treatment with the highly pyrophoric compound n-butyllithium. Furthermore, the conversion disrupts the semiconducting and photoluminescence properties of the pristine 2H phase. Lei et al. employed a Lewis acid-base mechanism to form coordination bonds with Se or S atoms on the TMDC surface using metal chlorides like $TiCl_4$ and $SnCl_4$.[32] $TiCl_4$, however, is highly volatile and reacts explosively with water releasing HCl, while $SnCl_4$ can decompose in air and was once employed as a chemical weapon.[33] Milder chemistries for the TMDCs have been reported, but they have been limited to interactions at defect sites or non-covalent bonds. Coordination chemistry using metal acetates[34] and ligand conjugation at defect sites in semiconducting TMDCs has been demonstrated,[35,36] but the maximum degree of surface functionalization is inherently limited to the initial concentration of defect sites. Noncovalent functionalization of TMDCs[37] has been demonstrated using various organic[13,23,35,38-40] and inorganic chemical species;[41] however, these methods rely only on physisorption to the semiconductor surface and are not as robust as covalent chemistries.

Consequently, there exists an unmet need for mild chemistries that will enable direct covalent functionalization of semiconducting TMDC surfaces. Here, we report the direct covalent functionalization of the basal plane of unmodified semiconducting $MoS_2$ using mild conditions and aryl diazonium salts, which can be used to covalently tether a wide variety of different functional groups to the $MoS_2$ surface. The chemical diversity that is available with the diazonium salts combined with their ready formation of covalent surface bonds thus establishes a versatile platform for chemical modification of semiconducting TMDCs. Our functionalization approach preserves the semiconducting and photoluminescence properties of $MoS_2$ without conversion to the metallic phase. The earlier reports[30,31] on $MoS_2$ functionalization using lithium pre-treatment to change the phase from semiconducting to metallic have widely spread the notion that this modification is a requirement.[29,42-44] However, we find that unmodified $MoS_2$ can indeed be covalently functionalized. We explore this finding using experiments and first principles calculations, and put forth a novel cooperative reaction mechanism that enables covalent functionalization of the highly inert semiconducting $MoS_2$ surface: the initial covalent attachment to a single sulfur vacancy enhances the reactivity at neighbouring lattice sites and enables the reaction to propagate rapidly across the otherwise pristine $MoS_2$ surface.

The formation of covalent C–S bonds is confirmed by X-ray photoemission spectroscopy (XPS), Fourier transform infrared spectroscopy (FTIR), and thermogravimetric analysis (TGA). Density functional theory (DFT) including van der Waals interactions is used to explain the reaction mechanism at the atomistic level. Scanning tunneling microscopy (STM) is used to image the $MoS_2$ atomic lattice, while atomic force microscopy (AFM) is used to characterize the morphology of molecular attachment as a function of reaction time. Raman and photoluminescence (PL) spectroscopic mapping are used to characterize the optical properties of covalently functionalized $MoS_2$, showing that the material remains semiconducting. Furthermore, we show that this approach is general by using it to effectively covalently functionalize $MoS_2$ obtained from three different methods: mechanical exfoliation, chemical vapour deposition, and solution phase dispersion. Lastly, we show that our technique can be extended to the covalent tethering of the active fluorescent proteins GFP and mCherry to the $MoS_2$ surface, demonstrating the utility of this approach toward future biological applications such as drug delivery,[45] bioimaging,[46] and biosensing.[27] This direct covalent modification of the $MoS_2$ surface suggests many future opportunities to enhance the properties of TMDCs via mild chemistries and provides a versatile chemical platform for tailoring the properties of $MoS_2$ simply by functionalizing with different aryl diazonium salts.

**Experimental Section**

*Mechanical exfoliation of $MoS_2$*

$SiO_2$ (300 nm)/Si substrates (Wafernet, Inc.) were ultrasonically cleaned in sequential baths of acetone and isopropanol and then blown dry with ultrahigh purity nitrogen gas. $MoS_2$ flakes were prepared on the cleaned substrate by mechanical exfoliation from a bulk $MoS_2$ crystal (SPI Supplies) using adhesive tape. The samples were then annealed in vacuum at 300°C to remove tape residue. Monolayer, bilayer, and few-layer flakes were identified by optical microscopy and Raman spectroscopy.

*Diazonium functionalization of $MoS_2$*

$MoS_2$ samples supported on $SiO_2$/Si substrates were immersed in 10 mM aqueous solutions of 4-nitrobenzenediazonium tetrafluoroborate (4-NBD) (Sigma Aldrich) with constant stirring (125 rpm) at 35°C for defined reaction times in a parafilm-sealed container kept in the dark. After each reaction step, the sample was gently rinsed with ultrapure water and blown dry with ultrahigh purity nitrogen gas before characterization to remove physisorbed molecules.

*Atomic force microscopy imaging*

AFM imaging was conducted using a Multimode V system (Bruker Corp.) with ScanAsyst-Air tips (Bruker) in ScanAsyst noncontact mode. Images were processed using the Gwyddion software package.[47]

*Raman and photoluminescence spectroscopy and mapping*

Raman and photoluminescence (PL) spectroscopies were performed in air at room temperature on a WITec alpha300R confocal Raman microscope system using a 532 nm excitation laser, 100X objective lens with ~1 μm diameter spot size. The laser power was kept between 0.26 to 0.32 mW to avoid damaging the MoS$_2$. The integration times were 1 s. Spatial maps of Raman and PL spectra were acquired at 30 pixels x 30 pixels, using the 300 grooves/mm grating for the PL spectra and the 1800 grooves/mm grating for the Raman spectra. The peak positions, total area intensities, and widths are obtained by Lorentzian fits of the spectra using Matlab. The error of the peak position from fitting is estimated to be ~0.5 cm$^{-1}$.

*Scanning tunneling microscopy (STM) imaging*

STM imaging of MoS$_2$ was conducted in an Omicron VT system operating at ultrahigh vacuum (UHV) conditions (base pressure 10$^{-10}$ mbar) and room temperature using electrochemically etched W tips. Images were processed using Gwyddion.[47] Samples were gently degassed at 200-300°C overnight in vacuum before imaging.

*Ar plasma treatment*

Point defects (mainly vacancies) in MoS$_2$ were generated by Ar plasma bombardment in a Plasmatherm 790 reactive ion etching (RIE) system with 20 sccm flow of Ar, 250 mtorr system pressure, 6 W setpoint power, and 4 s processing time. The actual power ranged from about 0.7 to 4.8 W during the 4 s of processing.

*van der Waals ab initio methods*

The calculations reported here are based on *ab initio* density functional theory using the SIESTA method[48] and the VASP code.[49,50] Results shown herein were produced using VASP, while SIESTA was used to perform initial tests with large numbers of atoms in the unit cell. The generalized gradient approximation[51] along with the DRSLL[52] functional, which includes vdW dispersion forces, were used in both methods, together with a double-ζ polarized basis set in SIESTA, and a well-converged plane-wave cutoff of 500 eV in VASP. Projected augmented wave method (PAW)[53,54] for the latter, and norm-conserving (NC) Troullier-Martins pseudopotentials[55] for the former, have been used in the description of the bonding environment for Mo, S, C, N, O, and H. The shape of the NAOs was automatically determined by the algorithms described by Soler et al.[48] The cutoff radii of the different orbitals were obtained using an energy shift of 50 meV, which proved to be sufficiently accurate to describe the geometries and the energetics. Atomic coordinates were allowed to relax until the forces on the ions were less than 0.01 eV/Å under the conjugate gradient algorithm. To model the system studied in the experiments, we created large supercells containing up to 212 atoms to simulate the functionalization between diazonium salts and the basal plane of MoS$_2$. To avoid any interactions between supercells in the non-periodic direction, a 20 Å vacuum space was used in all calculations. In this way, molecules are allowed to interact only along of the in-plane supercell directions, removing any artificial energy contribution from the out-of-plane periodic images. In addition to this, a cutoff energy of 120 Ry was used to resolve the real-space grid used to calculate the Hartree and exchange correlation contribution to the total energy. The Brillouin zone was sampled with a 11×11×1 grid under the Monkhorst-Pack scheme[56] to perform relaxations with and without van der Waals interactions. Energetics and electronic band structure were calculated using a converged 40×40×1 k-sampling for the unit cell of $MoS_2$-NP. In addition to this we used a Fermi-Dirac distribution with an electronic temperature of $k_BT$ = 20 meV to resolve the electronic structure.

## *Chemical vapour deposition (CVD) growth of $MoS_2$ for XPS*

Continuous centimeter-scale thin films of $MoS_2$ for x-ray photoelectron spectroscopy were grown on $SiO_2$/Si substrates that were first sonicated in sequential baths of acetone and isopropanol for 5 min each, blown dry with ultrahigh purity nitrogen gas, and cleaned in oxygen plasma (Harrick Plasma) at high power for 10 min. Solid powder precursors $MoO_3$ (15 mg, Sigma-Aldrich, 99%) and S (100 mg, Alfa Aesar, 99.5%) were loaded into separate quartz boats (MTI Corp.) and placed into a 1" diameter quartz tube in a hot-wall tube furnace (Lindberg). The $MoO_3$ boat was positioned in the center of the furnace with the target $SiO_2$/Si substrate placed face down across the upper edges of the boat. The S boat was positioned at the edge of the heating zone where the temperature reaches about 170°C during growth. The quartz tube was pumped down to ~6 mtorr vacuum before flowing 300 sccm of ultrahigh purity Ar gas, so that the chamber pressure was ~1.35 torr during growth. The furnace was heated from room temperature to 650°C over 40 min, kept at 650°C for 30 min, and then cooled rapidly by shutting off the furnace and then cooling with an external fan. This growth procedure results in relatively large area, continuous, and uniform coverage of monolayer and bilayer polycrystalline $MoS_2$. The samples are left on the $SiO_2$/Si substrates and directly used in the XPS measurements. The four diazonium-functionalized samples were reacted for different reaction times (10 s, 5 min, 10 min, and 6 h) and rinsed with ultrapure water and blown dry with ultrapure nitrogen gas before XPS measurement. The nitrobenzene control samples were dipped directly into undiluted nitrobenzene for 4 h and rinsed with isopropanol and dried before XPS measurement.

## *X-ray photoelectron spectroscopy (XPS)*

XPS spectra were acquired using a Vacuum Generators 220i-XL system with monochromated Al Kα radiation (hv = 1486.6 eV), linewidth 0.7 eV, spot size ~400 μm, and chamber pressure ~$10^{-9}$ torr or lower. Spectra were analyzed using the CasaXPS software package to subtract the Shirley backgrounds and fit the peaks to Gaussian/Lorentzian functions. Peak positions were shifted using the Si 2p peak from the substrate as a reference. Peaks were identified by comparison to known standards and the La Surface database from Centre national de la recherche scientifique (CNRS) in Orleans, France, and ThermoFisher Scientific (www.lasurface.com).

*MoS$_2$ dispersions and functionalization in solution*

MoS$_2$ was dispersed in sodium dodecyl sulfate (SDS) aqueous solution by microtip probe sonicating (Branson Digital Sonifier 450D, 13 mm diameter tip) 8.25 g of MoS$_2$ in 110 mL of 1% SDS solution (w/v) for 2 hours in a 250 mL steel beaker at 50% amplitude (power output of 48-50 W). Then 25 mL of this dispersion was transferred in 4 separate 50 mL plastic tubes and centrifuged at 4200 rpm for 3.5 hours to remove large, undispersed particles. To carry out the functionalization, 100 mg of the diazonium salt was added to 20 mL of the MoS$_2$-SDS dispersion and probe sonicated for 2 hours at 20% amplitude with a 3 mm diameter tip in a 50 mL tube. The resulting functionalized dispersion was flocculated with ethanol and filtered over a hydrophilic PTFE membrane (Omnipore, 100 nm pore size) and washed thoroughly with water and ethanol, resulting in a dry film collected on the membrane.

*FTIR and UV-vis characterization of bulk dispersions of MoS$_2$*

The dried films of functionalized MoS$_2$ on filter membrane were used to collect Fourier transform infrared (FTIR) spectra using a Nicolet 6700 system equipped with a Smart Orbit accessory. To re-suspend the samples in solution, the filter membranes were placed in 50 mL tubes along with 15 mL of SDS solution and bath sonicated for 2 hours. After sonication, the dispersions were filtered using Millipore vacuum filtration system (20 μm pore size) and then their UV-Vis absorbance spectra were collected (Jasco V760 UV-Visible/NIR Spectrophotometer). A control sample was also prepared in parallel using the same conditions as described above, except without the diazonium salt.

*Thermogravimetric analysis (TGA) of MoS$_2$ dispersions*

To prepare samples for TGA, 10 mL of the MoS$_2$ dispersion after 4-NBD functionalization was mixed with acetone in a ratio of 1:5 to aggregate and remove the SDS surfactant. After aggregation, the resulting dispersion was centrifuged for 30 minutes at 5000 rpm. The supernatant was decanted and the mixture was washed with 40 mL of DI water. The washing step was repeated three times. After washing, the sample was freeze dried to obtain a solid green powder, which was then analyzed using TGA. A control sample of the SDS-dispersed MoS$_2$ (without diazonium functionalization) was similarly processed to obtain solid green powder for TGA analysis.

TGA characterization was performed using a Setaram TG92 system. Each sample was purged with ultrahigh purity He gas overnight before TGA measurement. The He gas flow rate during the purge and the measurement was 30 mL per minute. The heating ramp rate was 5°C per minute up to 900°C. The first derivative curve (DTG) was calculated in Matlab by first smoothing the TG curve using a Savitzky-Golay filter and then taking the numerical derivative.

*SEM characterization of dried films from MoS$_2$ dispersions*

Dried films of functionalized and unfunctionalized MoS$_2$ on filter membranes were imaged by field emission SEM (Hitachi S4700 system) to characterize their morphologies.

*4-carboxybenzenediazonim tetrafluoroborate synthesis and characterization*

4-carboxybenzenediazonium tetrafluoroborate (4-CBD) was synthesized following a reported procedure.[57] Briefly, 1.35 g (0.01 mol) of p-aminobenzoic acid was dissolved in 14 ml of water and 3 ml of concentrated HCl. The mixture was cooled in an ice water bath until precipitates appeared. The precipitates disappeared after slow addition of sodium nitrite solution. The sodium nitrite solution was prepared by dissolving 0.752 g (0.011 mol) of sodium nitrite in 4 ml of water. The solution was vacuum filtered and then 1.465 g (0.013 mol) of sodium tetrafluoroborate was added. Then the solution was cooled below 0°C to obtain white crystals, which were then vacuum filtered and washed with ice-cold ether and water. The diazonium salt was dried and then stored at 4°C.

*Protein synthesis and purification*

The green fluorescent protein (GFP) expression plasmid was constructed following previously described methods using the GFP variant GFPmut3b.[58] The GFP gene was inserted into the pET15b (EMD Millipore) expression vector downstream of a T7 promoter and the polyhistidine tag sequence, yielding an N-terminal his-tagged GFP; and upstream of the T7 transcriptional terminator. The resulting plasmid was transformed into *E. coli* BL21 Star DE3. The transformed cells were cultured in 1 ml of LB medium with ampicillin (100 μg/ml) shaking at 37°C in an incubator overnight. The overnight culture was diluted 1:600 with fresh LB medium containing ampicillin (50 μg/ml) and grown until its absorbance at 600 nm reached 0.6-0.8. IPTG (isopropyl β-D-1-thiogalactopyranoside) was added into the culture to a final concentration of 0.5 mM to induce expression of T7 RNA polymerase and in turn trigger GFP production. After 4 hours of induction, the cells were harvested by centrifugation at 4000 *g* for 15 minutes. The cell pellet was resuspended in 27 ml lysis buffer (60 μg/ml lysozyme, 3.7 mM NaH$_2$PO$_4$, 16.3 mM Na$_2$HPO$_4$, 50 mM NaCl, 10 mM imidazole, 0.1 Protease Inhibitor Cocktail Tablet/ml) and sonicated at 4 W using a microtip probe (Branson Digital Sonifier 450D, 3 mm diameter tip). Three hundred 2-second pulses with a 2-second off time between each pulse were performed in an ice bath. The lysate was then centrifuged at 12,000 *g* for 30 min at 4°C. Approximately 25 ml of supernatant was collected and filtered through a 0.22-μm membrane filter. Purification was performed using fast protein liquid chromatography (FPLC) with a HisTrap HP column.

After equilibrating the column using 100% Buffer A (3.7 mM $NaH_2PO_4$, 16.3 mM $Na_2HPO_4$, 500 mM NaCl, 20 mM imidazole, 0.3 mM TCEP, pH 7.5), 10 mL cleared lysate was loaded into the column, washed with 10% of Buffer B (3.7 mM $NaH_2PO_4$, 16.3 mM $Na_2HPO_4$, 350mM NaCl, 500 mM imidazole, 0.3 mM TCEP, pH 7.5) and 90% of Buffer A, and eluted with 50% of Buffer B and Buffer A. Fractions were collected by monitoring the absorbance at 280 nm for the peak in its profile associated with the purified protein. Purified fractions collected from three FPLC runs were concentrated using Amicon (10 kD cutoff filters) and then stored in 50 mM $NaH_2PO_4$, 300 mM NaCl, pH 8.3. For smaller protein preparations, Ni-NTA spin columns (Qiagen) were used for purifying the His-tagged proteins. mCherry plasmid and protein preparation were performed using the same procedures as those used for GFP.

*Protein attachment*

The protocols used for protein attachment were similar to previous reports of protein attachment to graphene.[59,60] $MoS_2$ flakes exfoliated onto $SiO_2$/Si wafers were immersed in a 10 mM solution of 4-CBD and heated to 53-55°C for 2 hours. Then the samples were washed with water, acetone, IPA and water again sequentially. After drying, they were immersed in a solution of 2 mM EDC and 5 mM sulfo-NHS solution prepared in 2-(N-morpholino)ethanesulfonic acid (MES) buffer (0.1 M MES sodium salt, 0.5 M NaCl, pH adjusted to 6 with 1.0 N HCl) for 20 min. The samples were rinsed with water and immediately immersed into 11.3 mM solution of $N_\alpha,N_\alpha$-Bis(carboxymethyl)-L-lysine hydrate ($NTA-NH_2$) prepared in 1x PBS for 2 hr. The wafers were washed with water and dipped in 11.3 mM solution of $NiCl_2$ for 40 min. The wafers were again rinsed with water and immersed in 8 μM solutions of His-tagged GFP or His-tagged mCherry or 1:1 mixtures of the two proteins for 1 hour and rinsed with water twice and then air dried.

*Confocal microscopy imaging*

Confocal fluorescence microscope images were collected with a Leica TCS SP5 Spectral Confocal System using lasers with 488 nm and 561 nm wavelengths.

**Results and discussion**

*Chemical functionalization by aryl diazonium salts*

Aryl diazonium salts were used to functionalize $MoS_2$. These reagents have been previously used to covalently functionalize carbon nanotubes,[17,61] graphene,[16,60,62-65] and phosphorene[66]; however, these other materials have very different electronic structures and chemical reactivity than $MoS_2$, which has a highly inert surface. Unexpectedly, we found direct evidence of covalent functionalization of unmodified semiconducting MoS$_2$ by aryl diazonium salts. We first used 4-nitrobenzenediazonium (4-NBD) tetrafluoroborate salt in aqueous solution to functionalize MoS$_2$ (see Methods for more details). This diazonium salt is a model system that been used to functionalize other nanomaterials, and is chosen here so that we can focus on the unique reaction mechanism with MoS$_2$. A schematic illustration of the reaction process is shown in **Figures 1a-c**: (a) the 4-NBD diazonium salt in solution approaches the MoS$_2$ surface; (b) charge rearrangement at the surface allows the N$_2$ group to break off and form a nitrogen molecule while the remaining aryl group becomes a radical; and (c) a covalent C–S bond forms to produce MoS$_2$ functionalized by nitrophenyl (NP) groups.

The covalent functionalization was first demonstrated using mechanically exfoliated MoS$_2$ single crystal flakes. An optical microscope image of an atomically thin MoS$_2$ flake on a SiO$_2$/Si substrate is shown in **Figure 1d**. This flake was characterized by Raman spectroscopy and AFM imaging to identify the layer numbers of the marked regions as monolayer (1L), bilayer (2L), and four-layer (4L), with the height profile from AFM imaging along the dashed line shown in the inset of **Figure 1e**. (Additional Raman data are shown in **Figures S1 and S2** of the Supporting Information.) The typical photoluminescence (PL) spectrum of the 1L MoS$_2$ from this sample is shown in **Figure 1f**, with the A and B excitons and A$^-$ trion peaks labeled.[5,39,67] The spatial map of the total PL intensity in **Figure 1h** shows the higher PL from monolayer MoS$_2$ due to its direct bandgap.[5] After 5 s of functionalization by 4-NBD, the PL spectrum shows only small changes: a slight increase in intensity and very similar energies and peak shapes (**Figure 1g**), suggesting that the semiconducting electronic structure of MoS$_2$ is not disrupted by the chemical bonds. (Further analysis of the PL of MoS$_2$ with chemical functionalization is shown in the Supporting Information in **Figures S4 to S7**.)

Changes in the sample topography with 4-NBD functionalization are observed by AFM. In **Figures 1j-k**, AFM images show the region marked by the dotted square from **Figure 1e**. This area features a large 1L region of MoS$_2$, and a smaller 2L region on the left, which both initially appear smooth and flat. After 5 seconds of reaction with 4-NBD, **Figure 1k** shows numerous protrusions on the MoS$_2$ surface. These features have appeared after functionalization, and we interpret them to be organic functional groups attached to the surface, which is discussed in further detail below. The protrusions are arranged in a chain-like geometry and display no order in their attachment across the surface. A slightly higher density of protrusions is present on the 2L region compared to the 1L, indicating thickness dependent functionalization of the MoS$_2$, which is discussed in further detail below.

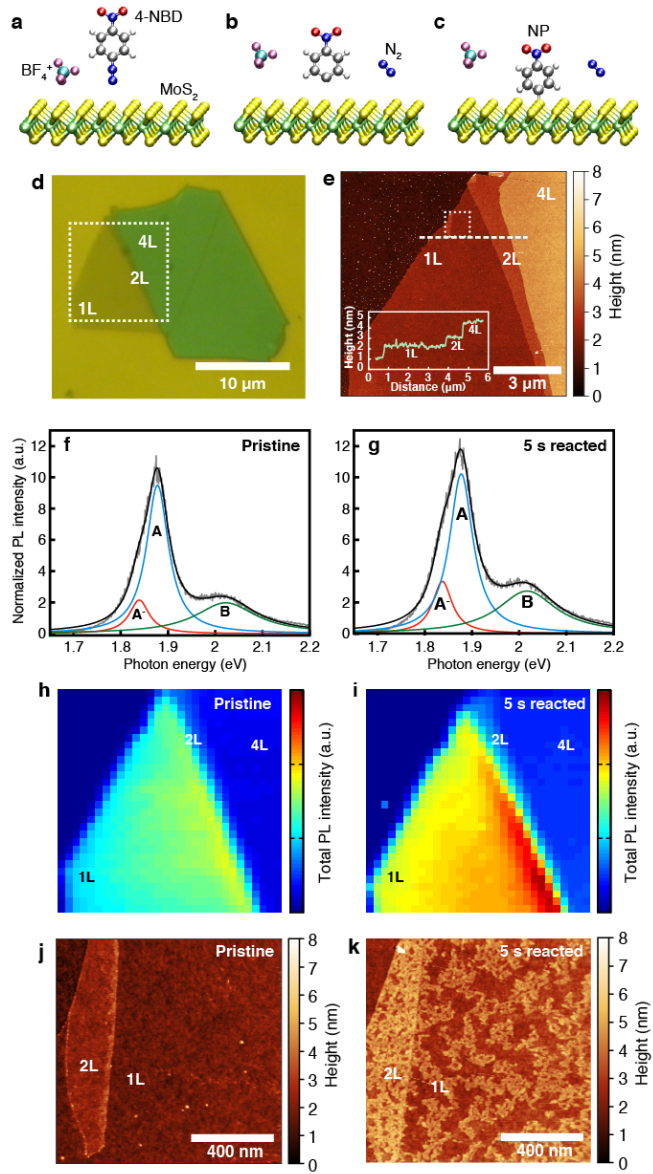

**Figure 1: Covalent functionalization of MoS$_2$. (a)-(c)** Schematic illustrations of the functionalization reaction. The aryl diazonium salt 4-nitrobenzenediazonium (4-NBD) tetrafluoroborate (BF$_4$) is dissolved in aqueous solution. Charges at the MoS$_2$ surface cause the diazonium group to break from the molecule as a nitrogen (N$_2$) molecule. The resulting nitrobenzene radical forms a covalent bond to a sulfur atom on the surface, resulting in a nitrophenyl (NP) functional group. **(d)** Optical microscope image of mechanically exfoliated MoS$_2$ flake with monolayer (1L), bilayer (2L), and four-layer (4L) regions marked. **(e)** Atomic force microscope (AFM) image of the region marked by the dotted square in panel (d) of the pristine MoS$_2$ flake. Inset: Height profile along the dashed line. **(f)** Representative photoluminescence (PL) spectrum of 1L MoS$_2$ before reaction with 4-NBD, and **(g)** after 5 s reaction. Lorentzian lineshapes were used to fit the A and B exciton and A$^-$ trion peaks. The spectra are normalized to the height of the Raman peaks. **(h)** Spatial map of total integrated intensity of PL for MoS$_2$ in the region marked by the dotted square in panel (d) before reaction and **(i)** after 5 s reaction. The PL intensity is highest for monolayer MoS$_2$ due to its direct bandgap. **(j)** AFM image of the region in the dashed square in panel (e), with mainly 1L MoS$_2$ and a small region of 2L MoS$_2$, before reaction and **(f)** after 5 s reaction. Many small protrusions in chain-like shapes are observed.

*Atomistic mechanisms of chemical functionalization*

To better understand the chemical functionalization of $MoS_2$ by diazonium molecules, we performed density functional theory (DFT) calculations including van der Waals (vdW) dispersion forces as shown in **Figure 2**. (See Methods section for details.) Overall, we obtain strong theoretical evidence for a cooperative reaction mechanism where the existence of a single initiation site allows the rest of the pristine semiconducting $MoS_2$ surface to be covalently functionalized. The simulations show that the covalent functionalization reaction with 4-NBD indeed cannot proceed for completely pristine $MoS_2$ due to its unfavorable energetics (see reactions **IV** and **V** in **Figure S8** of the Supporting Information). Instead, the presence of a single sulfur vacancy (S-vacancy) allows the reaction to proceed with a stable binding energy (from -1.03 to -1.53 eV per molecule), which is downhill in energy as the chemical reaction evolves. This suggests that S-vacancies serve as nucleation centers for the initial functionalization. (See **Figure S8** of the Supporting Information for additional scenarios between the 4-NBD molecule and $MoS_2$ surface that were considered, and **Figure S9** for the energetics of the reaction as it proceeds.)

The role of covalently attached molecules in the overall reaction mechanism was also explored by DFT, leading to a picture of a cooperative process. **Figure 2a** shows the calculated binding energy per 4-NBD molecule as a function of molecular coverage levels for different S-vacancy concentrations. In the pristine $MoS_2$ case (0.00% concentration of S-vacancies), once the first NP groups bind to the surface, they induce charge reorganization and increased reactivity in the surrounding $MoS_2$ in an area in the range of 9.23 to 36.92 $Å^2$ (see **Figure S10** in Supporting Information). This change decreases the energetic barrier for the adsorption and reaction of subsequent 4-NBD molecules, so the binding energies increase in magnitude (become more negative) as a function of increasing molecular coverage. We observe that this effect extends for several unit cells across the $MoS_2$ surface, and that it is specific to S-atoms in the same sublattice (**Figure 2b**). This result indicates that the molecule itself plays an active role in propagating the surface functionalization, creating new nucleation centers for neighbouring molecules even if there are no initial $MoS_2$ defects. In this process, the molecules tend to be packed together forming chains as indicated in the calculated molecular configurations in **Figure 2b** and the schematic in **Figure 3a**, rather than attaching at isolated and disperse locations. That is, the molecules are behaving cooperatively during the functionalization process to help propagate the reaction. This behavior is an important finding in the overall covalent functionalization mechanism.

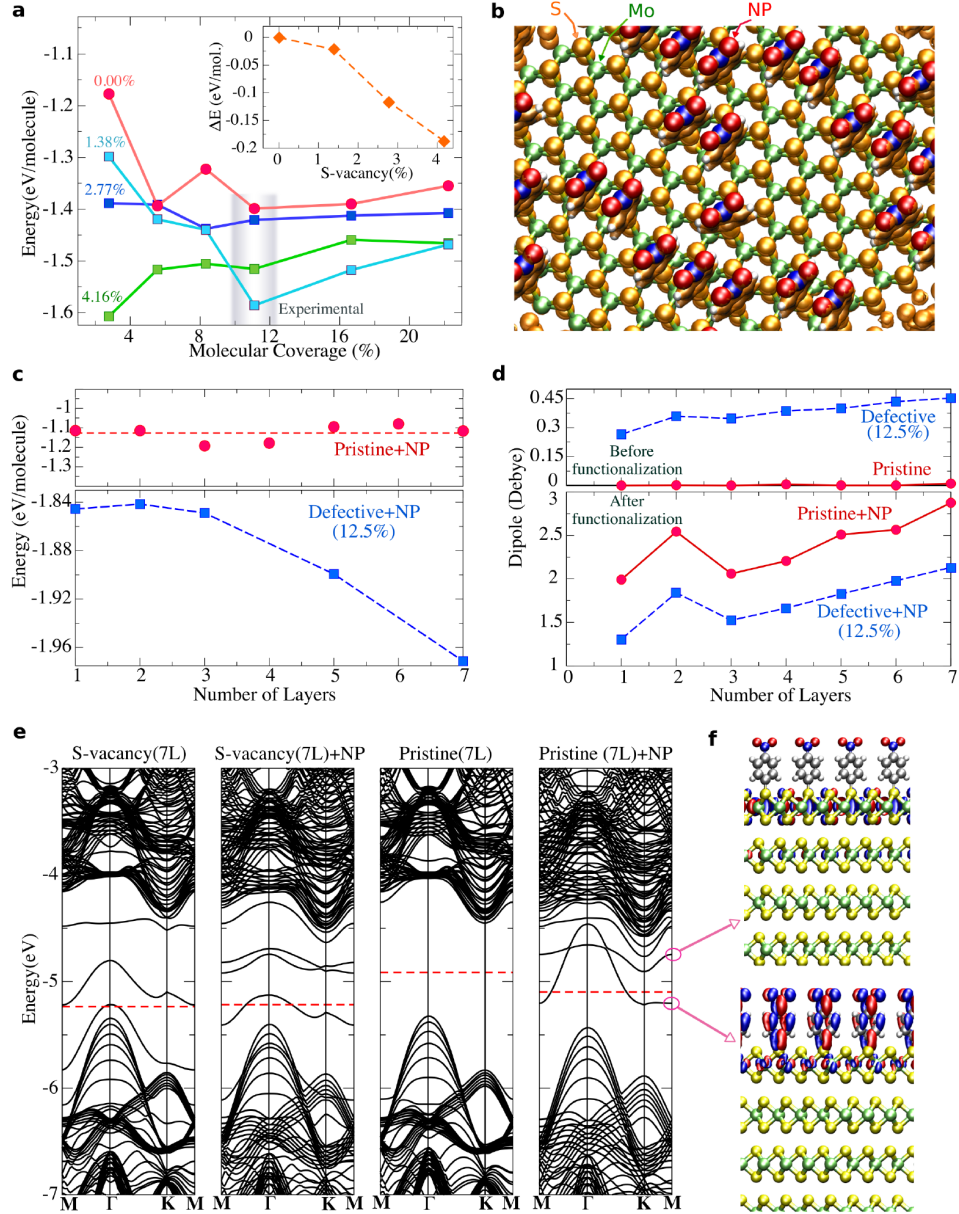

**Figure 2: Theoretical calculations of the effects of defects, coverage and number of layers on the functionalization of MoS$_2$.** **(a)** Binding energy (eV) per molecule as a function of molecular coverage (%) on the surface of monolayer MoS$_2$. The binding energy is defined as E = E[MoS$_2$+NP] − E[NP] − E[MoS$_2$]. Different S-vacancy concentrations ranging from 0 to 4.16% are shown as coloured traces. The shaded region indicates the estimated concentration of reacted groups from our experiments. The inset shows the variation of the binding energy relative to pristine surface as S-vacancies are introduced into the system. **(b)** Isosurface of the charge density distribution (±0.8 e/Bohrs$^3$) of NP molecules on MoS$_2$ surface at no defects. Molecular coverage of 11.1%. Color scheme shows: Mo (green), S (underneath the charge density in orange), O (red), and N (blue). **(c)** Binding energy (eV per molecule) as a function of the number of layers for pristine (upper panel) and defective (lower panel) MoS$_2$ surface with 12.5% S-vacancy concentration. The dashed line in the pristine case is the average along the entire set of thicknesses, and in the defective case is the interpolation between the points. **(d)** Dipole moment at the MoS$_2$ surface as a function of the number of layers for pristine and defective (12.5% S-vacancies) MoS$_2$ before (upper panel) and after (lower panel) the functionalization. **(e)** Band structure of 7L pristine and defective (12.5% S-vacancy concentration) MoS$_2$ surface, before and after the functionalization takes place. The Fermi level is marked by the red dashed line in each panel. **(f)** Isosurface (±0.04 e/Bohrs$^3$) of the defect states near the Fermi level for 7L pristine MoS$_2$. Blue and red isosurfaces represent positive and negative values, respectively, of the Kohn-Sham orbitals at the M-point.

As the molecular coverage increases, a further increase in the stability is observed reaching some saturation at ~16%. Interestingly, when more molecules are added on the surface, slight decreases of the magnitude of the binding energies are observed, as some molecules can be detached from the S sites at the limit of ~50% coverage or when the intermolecular distance is about one unit cell (see **Movies S1** and **S2** in Supporting Information). This change in binding energies as a function of molecular coverage indicates a subtle balance between adsorbate–adsorbate and adsorbate–surface interactions, which drives the system to an optimal coverage. The process of NP groups detaching from the $MoS_2$ surface is observed to depend on the way that the benzene ring of the NP molecules interacts with their nearest neighbour molecules. The degree of torsion of the benzene ring determines the relative stability when several molecules are attached to the surface and interact through their $p_z$-orbitals (see **Movies S2** and **S3** and **Figures S10 and S11** in the Supporting Information). We observe a dip in the binding energies at several S-vacancy levels centred around 10% coverage and extending to about 8-12%, suggesting that there might be an optimal concentration. This range is close to the experimental coverage we estimate from thermogravimetric analysis (TGA) measurements discussed below, which is marked as the shaded range in **Figure 2a**.

Once defects such as S-vacancies are present in the top layer of $MoS_2$, we calculate that there is a further energy gain for attaching additional molecules. The dark blue, light blue, and green plots in **Figure 2a** show the binding energy for three different initial concentrations of S-vacancies as a function of molecular coverage. These results show that the stability of the NP molecules increases with increasing vacancy concentration. Using a molecular coverage of 11.2%, which is within our estimated optimum concentration, and varying the concentration of initial S-vacancies on the $MoS_2$ surface between 0 and 4% (see inset in **Figure 2a**), energy gains of 0.2 eV per molecule can be achieved at a small amount of disorder. This suggests that the addition of S-vacancies in $MoS_2$ can increase the initial rate of 4-NBD functionalization, but does not control the final concentration of functionalized sites.

Next, we investigated the effect of the number of layers on the energies associated with functionalization of pristine $MoS_2$ and $MoS_2$ with some defects, as experimentally observed in **Figure 1k**. Based on DFT calculations, **Figure 2c** shows that when no defects are present, there are small fluctuations in the binding energies (<0.05 eV) of NP on $MoS_2$ as a function of layer number, and that they are centred around the average value of about 1.12 eV per molecule (horizontal dashed line). Once S-vacancies are introduced at ~12.5% concentration, a systematic increase of the binding energy with layer number is observed with an energy gain of about 0.14 eV per molecule at 7 layers. This indicates that for $MoS_2$ with some defects, multilayers are more reactive than thin layers. This trend is consistent with our experimental observations that $MoS_2$ samples with more layers have a denser spatial coverage of covalently attached organic groups, as seen in AFM images of 1L, 2L and 4L regions in **Figures 1 and 3**, and in **Figure S13** of the Supporting Information.

The driving force for this layer- and defect-dependent effect can be appreciated in **Figure 2d**, which shows the variation of the surface dipole moment before (top panel) and after functionalization (bottom panel) as a function of layer number. In the absence of S-vacancies, no dipole moment is created at the top layer and the NP molecule interacts with the same bare surface regardless of the number of layers underneath. At a finite concentration of defects, a surface dipole moment is created due to the dangling bonds at the defect site (see **Figure S12** in Supporting Information) and increases slightly in magnitude with increasing layer thickness following the enhancement of polarization charge at the surface. This dipole makes the interactions with the NP molecules much more efficient and enhances the binding energy with the surface for different numbers of layers. This layer number dependent trend is in general agreement with the experimental results as shown in **Figures 1 and 3** for 1L, 2L and 4L $MoS_2$ layers.

It is worth noting that the 4-NBD molecule itself has a dipole moment as high as 3.83 Debye in the gas phase, which is partially present at either at the pristine or defective surface once the functionalization takes place (bottom panel in **Figure 2d**). This suggests that dipole-dipole interactions between the S-vacancy site and the 4-NBD molecule are important factors to catalyze the formation of the covalent bond on the surface. We can also observe that defect states become pinned at the Fermi level (**Figure 2e** and **Figure S12**), generating partially filled states that can create an additional charge density to stabilize the S-NP bond. Such defect states are observed at both the defective and pristine surfaces showing a character that is composed of a combination of the molecular orbitals from both the molecule ($2sp_z$ C-states) and the surface ($3sp$ S-states) (**Figure 2f**). For the defective surface, however, these states are already present before the functionalization, which explains the increase of the adsorption strength on the S-vacancy site relative to pristine surfaces. The strategy of introducing S-vacancies to initialize the functionalization of NP is therefore a contributing factor in the engineering of the electronic structure of the bare $MoS_2$ surface, in addition to the interaction between adsorbates.

*Surface coverage and initialization of functionalization*

Based on the DFT results described above, we propose the reaction scheme illustrated in **Figure 3a**. While the completely pristine $MoS_2$ basal plane is unreactive (light green area), the region around a S-vacancy or defect (white circle) has a higher density of states and thus enhanced reactivity (dark green area). Consequently, 4-NBD will react in that region of enhanced reactivity, resulting in a covalently bound NP group at a S atom. The newly attached NP group in turn establishes around it a new region of higher density of states and charge density, which contains contributions from both the overlapping charge densities of the functionalized site and the adsorption-induced charge reorganization from the new molecule. This charge rearrangement minimizes the Pauli repulsion, allowing for increasing electrostatic attraction between the basal plane and the NP group. This new region of enhanced reactivity then encourages attachment of additional molecules to the surface. Subsequent molecules that attach near the bound NP will further extend and reinforce the region of higher reactivity, thus resulting in chain-like growth propagation that can start from a single defect site and eventually span the entire MoS$_2$ surface. The intentional formation of S-vacancies by methods such as Ar plasma bombardment can thus be used to tailor the number of initiation sites for the reaction and alter the speed and paths over which the reaction propagates over the MoS$_2$ surface. Notably, this new reaction mechanism involving defects and propagating regions of higher charge density is different from models previously proposed for MoS$_2$ edges where a substantial charge density is present at the uncoordinated atoms.[35,68]

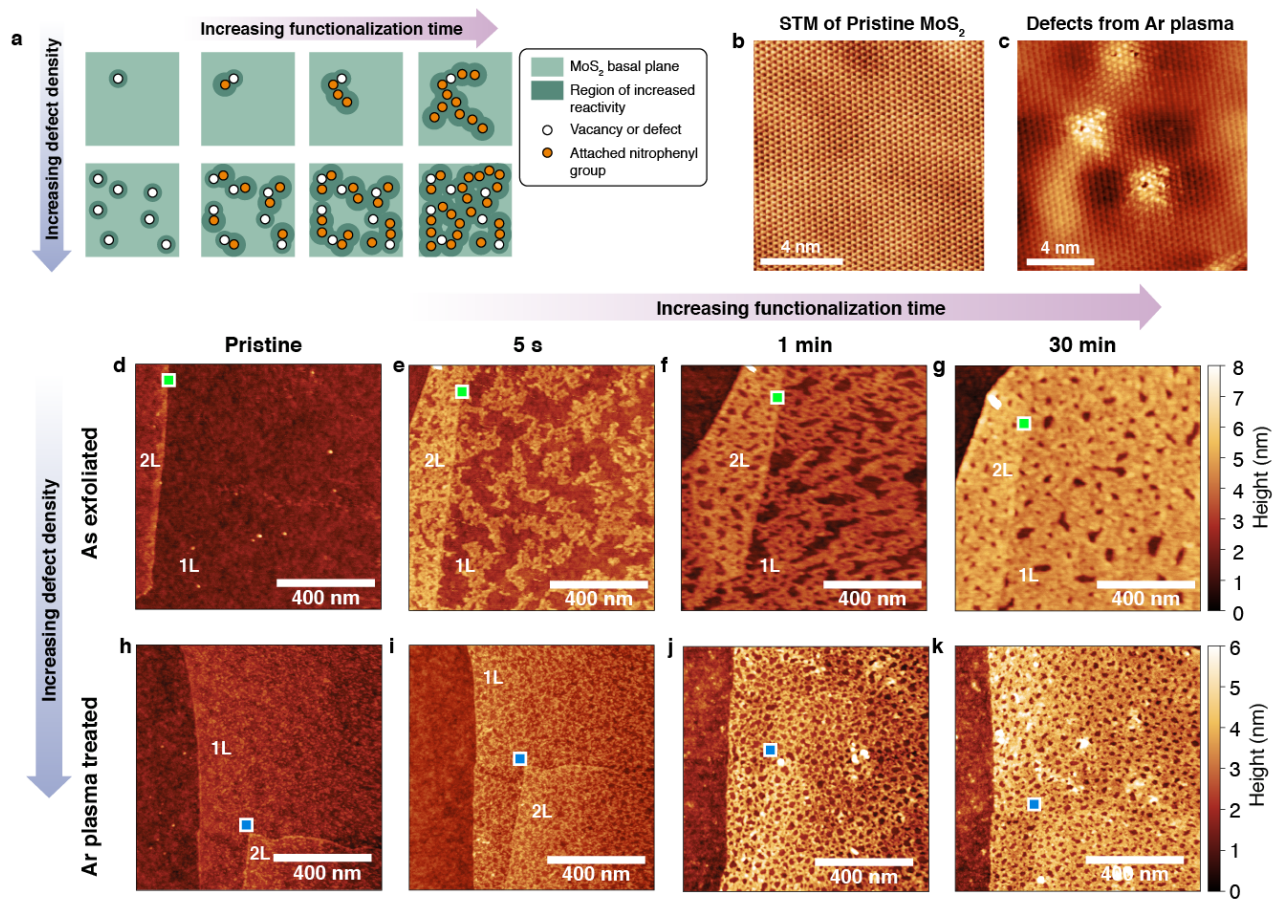

**Figure 3: Molecular coverage as a function of reaction time and initial defect sites.** (a) Schematic illustration of the growth of molecular coverage on MoS$_2$ with 4-NBD reaction time. The light green background represents the MoS$_2$ surface, the dark green represents regions of increased reactivity surrounding defects and covalently attached nitrophenyl (NP) groups, the white circles represent vacancies or defects, and the orange circles represent covalently attached NP groups. Molecules preferentially attach at regions of increased reactivity near defects or other previously attached molecules. (b) Scanning tunneling microscopy (STM) image of as-exfoliated atomic lattice of MoS$_2$. (Imaging conditions: -1.6 V sample bias, 0.4 nA tunneling current) (c) STM image of MoS$_2$ after Ar plasma treatment at 6 W for 4 s, showing point defects. (Imaging conditions: -1.7 V, 0.5 nA) (d)-(g) AFM images of as-exfoliated MoS$_2$ and after different 4-NBD reaction times: 5 s, 1 min, and 30 min. The green squares indicate the same location on the sample in each image. (h)-(k) AFM images of Ar plasma treated MoS$_2$ after different 4-NBD reaction times. The blue squares indicate the same location on the sample in each image. In all AFM images, 1L and 2L refer to monolayer and bilayer regions, respectively.

To further explore the effects calculated by DFT and to validate our model, we have carried out detailed scanning probe microscopy studies of functionalization of MoS$_2$ samples with and without initial defects and over different 4-NBD reaction times. On the pristine mechanically exfoliated MoS$_2$ samples, scanning tunneling microscopy (STM) imaging shows the atomic lattice without any defects (**Figure 3b**). In general, we observed very few defects while imaging over large regions of the sample. Previous work in the literature shows that S-vacancies are the most common type of defects in mechanically exfoliated MoS$_2$.[69] We then used Ar plasma treatment in a reactive ion etching (RIE) system to generate point defects in MoS$_2$ (see Methods for details).[70,71] STM images of the Ar plasma treated MoS$_2$ (**Figure 3c** and **Figure S14** in the Supporting Information) indicate an average defect concentration of ~1-2%, with some smaller areas being more defective (~4-5%). These defects appear to be small depressions surrounded by a ~1 nm radius of increased brightness, which we interpret to be S-vacancies.[72]

AFM imaging was used to track the progress of the functionalization reaction over time for both pristine and Ar plasma treated MoS$_2$ samples. For the as-exfoliated pristine sample (**Figure 3d**), at 5 s reaction time (**Figure 3e**) we observe small chain-like protrusions of ~2 nm height across the MoS$_2$ surface, with a higher density of coverage in the 2L region of the sample. There are no protrusions on the surrounding SiO$_2$ substrate. We interpret these protrusions to be covalently attached NP groups, as they are consistent with our previous experimental and DFT results, and because the sample is thoroughly rinsed at each reaction step to remove any physisorbed molecules. At 1 min reaction time (**Figure 3f**), the density of protrusions is higher, and the regions of unreacted MoS$_2$ are smaller. The initial chain-like clusters appear to have grown larger or longer, rather than new clusters nucleating. By 30 min reaction time (**Figure 3g**), the reacted groups are even more densely covering the MoS$_2$, and the remaining unreacted areas of MoS$_2$ are observed as small depressions between the molecular groups. We note that the denser molecular coverage in thicker 2L MoS$_2$ regions is also seen in the additional AFM images in **Figure S13** of the Supporting Information, which shows that there is higher coverage in 2L and 4L regions, in agreement with our DFT results for higher reactivity with increasing layer number shown in **Figure 2**.

The Ar plasma-treated MoS$_2$ sample is shown in **Figures 3h-k** for different 4-NBD functionalization times. Raman and PL spectra of the same sample (**Figure S3** and **S4** of the Supporting Information) show that the overall structure and electronic and optical properties of the MoS$_2$ are not significantly changed by the Ar plasma treatment. Upon functionalization with 4-NBD, the protrusions at 5 s reaction time (**Figure 3i**) are present at a much higher areal density than observed with the as-exfoliated MoS$_2$ (**Figure 3e**). With increasing reaction time the protrusions become interconnected on the surface, while still leaving some portions of MoS$_2$ unreacted (**Figure 3j-k**). These experimental observations of more reacted sites at short reaction times for the sample with more S-vacancy sites corroborate the picture of 4-NBD functionalization we have developed from DFT calculations.

*Optical and chemical characterization*

The preservation of the structure and semiconducting properties of MoS$_2$ with diazonium functionalization is shown by optical characterization using Raman and PL spectroscopies (see **Figures S4** to **S7** in the Supporting Information for more discussion). The Raman spectra of initially pristine MoS$_2$ and Ar plasma treated MoS$_2$ at increasing 4-NBD functionalization times show that the main A$_{1g}$ and E$^1_{2g}$ peaks remain unchanged, and that defect-associated peaks do not make a noticeable appearance. The PL spectra similarly show only small changes in positions and intensities of the main peaks, indicating that the electronic structure of the semiconducting MoS$_2$ is intact. In the initially pristine MoS$_2$ sample, the relative decrease in intensity of the exciton (A peak) with respect to the trion (A$^-$ peak) as well as the widening gap in positions between these two peaks with increasing reaction are both consistent with increasing n-doping (see **Figure S6** in the Supporting Information). Based on changes in PL peak positions, which are correlated with changes in exciton energies with doping, we estimate that the MoS$_2$ becomes more n-doped with ~7 meV increase in Fermi level (see Figure S6 and pages 7-8 in the Supporting Information for more details).[67]

X-ray photoelectron spectroscopy (XPS) was used to characterize the chemical bonding occurring from diazonium functionalization (see Methods section for more details). Large-area samples of polycrystalline MoS$_2$ grown by chemical vapor deposition (CVD) with a mixture of monolayer and bilayer regions were functionalized by 4-NBD under the same reaction conditions as shown in the previous experiments for different reaction times: 10 s, 5 min, 10 min, and 6 h. Control experiments were also conducted by immersing MoS$_2$ in concentrated nitrobenzene for 4 h. Nitrobenzene lacks the diazonium group that is crucial to the covalent reaction mechanism, so only physisorption is possible.[73] (See Supporting Information, **Figure S15**, for more data from nitrobenzene control experiments including AFM images that show the morphology of the physisorbed molecules on MoS$_2$ is quite different from the covalently bound molecules.)

The resulting XPS spectra from these experiments are shown in **Figure 4**, and are normalized to the intensity of the Mo peak since the Mo atoms are sandwiched between S atoms and do not participate in surface reactions. The black curves are the experimental data and the coloured curves are peak fits. (Further characterization of the CVD-synthesized MoS$_2$ is shown in the Supporting Information, **Figure S16**). We examine changes in the N, C, Mo, and S curves to obtain evidence that the 4-NBD reaction results in covalent functionalization and the growth of an organic layer, and that the nitrobenzene control lacks such evidence.

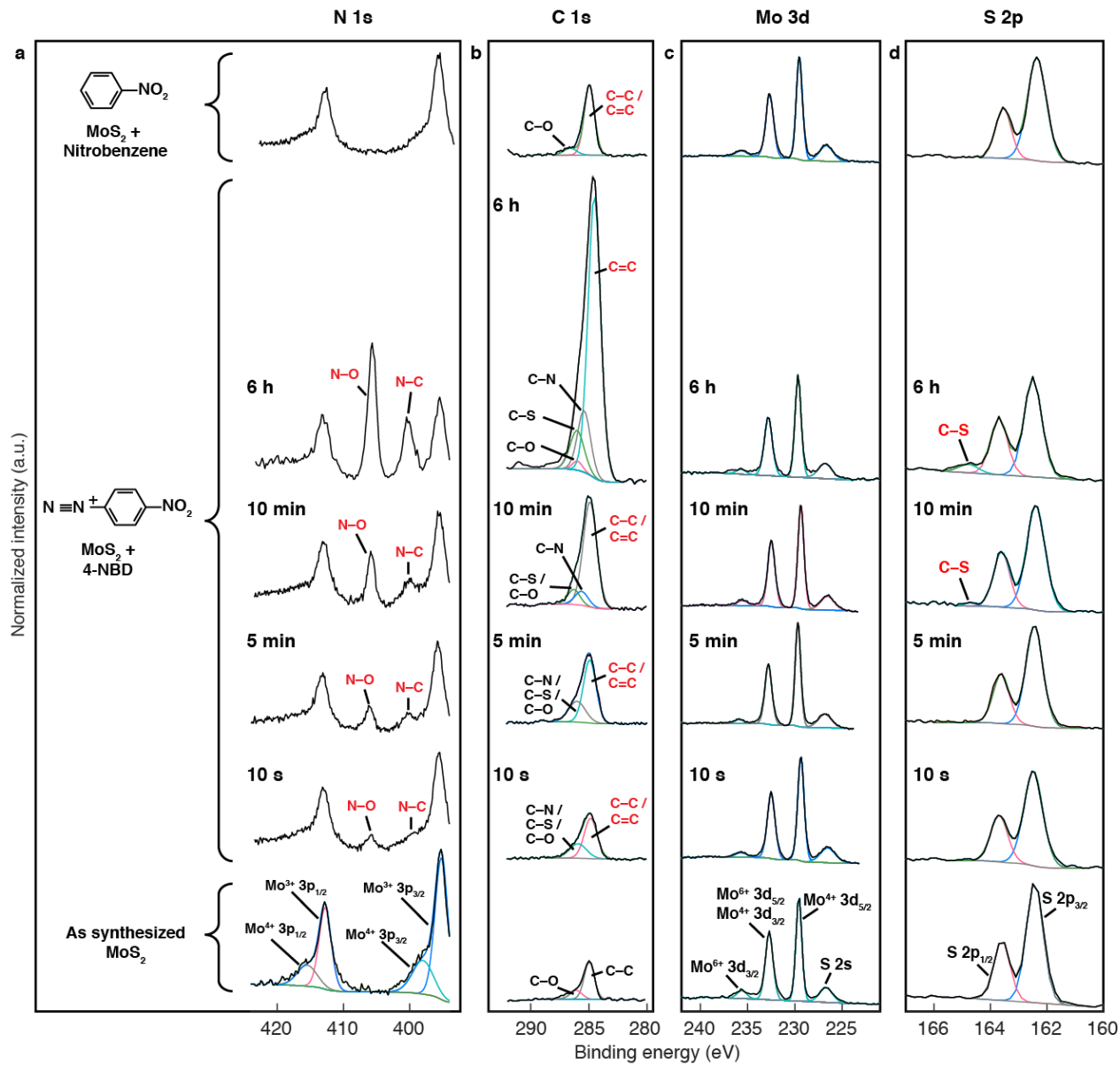

**Figure 4: X-ray photoelectron spectroscopy (XPS) of functionalized MoS$_2$.** XPS spectra for MoS$_2$ synthesized by chemical vapour deposition (CVD) (bottom spectrum), and reacted with 4-NBD for 10 s, 5 min, 10 min, and 6 h (middle four spectra). XPS spectrum of MoS$_2$ exposed to nitrobenzene as a control reaction (top spectrum) without the diazonium functional group, so that covalent functionalization is not possible and only physisorption is possible. Spectra are vertically offset for clarity. **(a)** Nitrogen 1s region: As-synthesized MoS$_2$ shows peaks from Mo$^{3+}$ and Mo$^{4+}$. After 4-NBD reaction, there are new N–O and N–C peaks, which increase in intensity with reaction time. The nitrobenzene control does not have these N-related peaks. **(b)** Carbon 1s region: As-synthesized MoS$_2$ shows C peak from adventitious carbon. After 4-NBD functionalization there is a peak from C=C in the benzene ring, and after nitrobenzene exposure the C peak is only slightly increased due to some residual physisorbed nitrobenzene. **(c)** Molybdenum 3d region: All spectra show nearly identical peaks corresponding to Mo and S. The Mo peaks have contributions from mainly MoS$_2$ but also some from small amounts of the MoO$_3$ precursor used in the CVD process. **(d)** Sulfur 2p region: all three spectra show the characteristic split peaks for S due to spin-orbit splitting. In the 4-NBD functionalized spectra at 10 min and 6 h, there is a new peak attributed to C–S bonds that increases in intensity with increasing reaction time. Wide scan XPS spectra are shown in Figure S17 of the Supporting Information.

In **Figure 4a**, we observe two peaks due to Mo 3p for all the reaction conditions. With increasing 4-NBD functionalization time, peaks from N–O and N–C bonds arise and increase in intensity, due to the $NO_2$ group in the 4-NBD molecule being attached to the $MoS_2$ surface. However, the nitrobenzene control sample shows no N peak, even though concentrated nitrobenzene was in contact with the $MoS_2$ for 4 h, indicating that no covalent bonds to the surface have formed; even the 10 s diazonium functionalized sample has a much larger N–O peak.

In **Figure 4b**, a small adventitious carbon peak occurs for as-synthesized $MoS_2$. With increasing 4-NBD functionalization time, the C peak attributed to aromatic C=C bonds, C–N bonds from the nitrophenyl groups[74] covalently attached to the $MoS_2$ surface, and C–S bonds at the functionalization sites increase in intensity and becomes quite large at 6 h, with the main contribution coming from C=C. In the spectrum for the nitrobenzene control, the C peak is similar in size to the initial adventitious carbon peak from unfunctionalized $MoS_2$, and perhaps slightly larger due to some residual physisorption of molecules. This behaviour is again consistent with no covalent bonding to the $MoS_2$ surface when the diazonium group is absent. There is also no C–Mo bond, indicating the NP molecules are only bonding with the S atoms at the top surface, and not with any Mo atoms in the middle of the S-Mo-S sandwich-like structure of $MoS_2$. In **Figure 4c**, the typical peaks associated with the Mo and S in $MoS_2$ and some small Mo peaks from with residual $MoO_3$ precursor[75,76] are not significantly changed after functionalization, since the Mo atoms do not participate in surface reactions.

In **Figure 4d**, we observe the two characteristic S peaks due to spin-orbit splitting. With 4-NBD functionalization, a new peak associated with C–S bonds[77,78] appears, suggesting the successful formation of covalent bonds between the C atoms in aryl groups and the S atoms at the top surface of $MoS_2$. This peak is relatively small, even for the 6 h diazonium functionalized sample, because not all surface atoms are reacted according to our DFT calculations, and because the samples are bilayer in some regions there is a larger contribution from the unreacted S below the top surface. From additional thermogravimetric analysis (TGA) measurements discussed below, we estimate ~12% coverage of covalently reacted sites on the $MoS_2$ surface, which is close agreement with DFT predictions described above. Another possible factor is that some of the 4-NBD groups may be attaching to existing covalently bound groups to form oligomers rather than to the bare $MoS_2$ surface, so that the number of C–S bonds does not increase in proportion to the overall amount of additional N and C on the surface with increasing reaction time. Similar oligomer formation was reported for aryl diazonium functionalization of graphene,[62] and the formation of bonds to existing molecules is generally more energetically favourable than the bond to the S-atoms by ~3-4 eV.

*Functionalization of solution phase dispersions of MoS$_2$*

To demonstrate the wide-ranging application of our functionalization method, we also functionalized bulk unmodified MoS$_2$ dispersed in aqueous surfactant solution (**Figure 5**). The resulting functionalized MoS$_2$ flakes provide further evidence of covalent bond formation and display enhanced stability in aqueous dispersions. The process of dispersion and functionalization is illustrated in **Figure 5a**. Bulk powder MoS$_2$ was probe-sonicated in an aqueous solution of sodium dodecyl sulfate (SDS) to exfoliate it into sheets that are stabilized by the SDS molecules (**Figure 5b(i)**). Then, 4-NBD was added and the mixture was probe-sonicated to allow the diazonium salt and pristine MoS$_2$ sheets to react. Sonication helped in producing more accessible sites between layers for the diazonium salt to react, encouraging more efficient functionalization. To remove excess 4-NBD, the resulting dispersion was flocculated with ethanol, collected on a filter membrane and washed thoroughly with water and ethanol. (See Methods section for additional details.) A control sample of unfunctionalized MoS$_2$ was also prepared following the same procedure, but leaving out the 4-NBD functionalization step. The unfunctionalized MoS$_2$ and functionalized 4-NBD/MoS$_2$ materials collected on the filter membranes form a continuous thick film with similar colours as their dispersions (see Figure S23 in Supporting Information).

The 4-NBD/MoS$_2$ and MoS$_2$ materials collected on the filter membranes were characterized by Fourier transform infrared (FTIR) spectroscopy. In **Figure 5c**, the FTIR spectrum of diazonium-functionalized MoS$_2$ clearly shows the presence of characteristic peaks that confirmed successful covalent modification MoS$_2$ in bulk dispersions. The peaks at ~1518 cm$^{-1}$ and ~1344 cm$^{-1}$ represent the stretching vibrations of the N-O bond in the NO$_2$ group, the peak at 1595 cm$^{-1}$ represents C=C stretching vibrations in the aromatic ring, and the peak at 697 cm$^{-1}$ can be assigned to S-C stretching vibrations at the covalent bond between the MoS$_2$ surface.

After acquiring the FTIR spectra, the films were re-dispersed in a fresh SDS solution using bath sonication (see Methods section for further details). The diazonium-functionalized MoS$_2$ material gave a highly concentrated dispersion after bath sonication (**Figure 5b(iii)**) in contrast to unfunctionalized MoS$_2$, which was only weakly re-dispersed (**Figure 5b(ii)**). This change in dispersibility can be attributed to a change in surface energy, a phenomenon which has been previously reported for dispersions of TMDCs,[34,79] and for covalently functionalized graphene.[19,20] The change in dispersibility is also reflected in the morphologies of the unfunctionalized and functionalized MoS$_2$ films, as shown in scanning electron microscopy (SEM) images in Figure S23 of the Supporting Information. The unfunctionalized MoS$_2$ forms a densely packed film that has many large cracks of 50-100 μm length and 5-10 μm width, but not many distinct particles or flakes. In contrast, the functionalized 4-NBD/MoS$_2$ film is much more uniform without any cracks, but more visible particles.

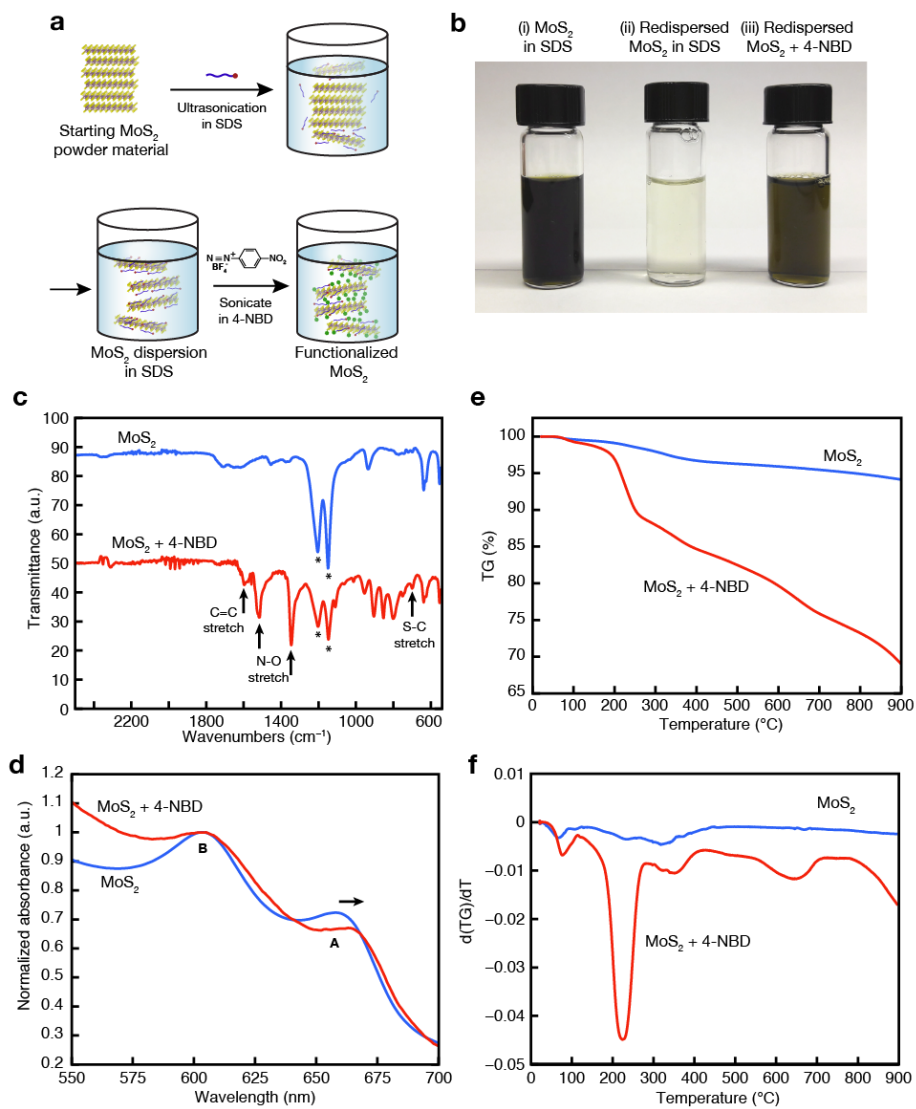

**Figure 5: Bulk solution phase functionalization of MoS$_2$.** **(a)** Schematic diagram of solution phase exfoliation and functionalization processes. The initial MoS$_2$ powder material was exfoliated and dispersed by ultrasonication in an aqueous solution of sodium dodecyl sulfate (SDS), which causes thin flakes of MoS$_2$ to separate and become encapsulated by SDS molecules. The dispersed MoS$_2$ flakes were then directly functionalized by 4-NBD in solution, resulting in a stable dispersion of covalently functionalized MoS$_2$. **(b)** Photographs of different MoS$_2$ preparations: **(i)** a dark green suspension of MoS$_2$ in SDS; **(ii)** a pale green suspension of MoS$_2$ that has been dispersed, dried and restacked, and redispersed in SDS, showing ineffective redispersion after restacking; and **(iii)** a dark green suspension of MoS$_2$ that has been dispersed and functionalized in 4-NBD, dried and restacked, and redispersed, showing that the functionalization allows for effective redispersion. All suspensions were centrifuged to remove undispersed parts. **(c)** Fourier transform infrared (FTIR) spectra for dried films of unfunctionalized MoS$_2$ (blue curve) and 4-NBD functionalized MoS$_2$ (red curve). Peaks corresponding to the nitrophenyl groups (C=C stretch, N-O stretch) and the covalent functionalization (S-C stretch) are indicated. The peaks marked (*) come from the PTFE substrate supporting the thin films. **(d)** UV-vis absorbance curves for MoS$_2$ suspensions in SDS (blue curve) and 4-NBD functionalized MoS$_2$ (red curve). The A and B exciton peaks are labeled. The A peak redshifts after 4-NBD functionalization. **(e)** Thermogravimetric analysis (TGA) mass loss curve for MoS$_2$ (blue curve) and 4-NBD functionalized MoS$_2$ (red curve). **(f)** TGA derivative (DTG) curves for the mass loss curves in panel (e), showing dips corresponding to key mass loss components.

Optical absorbance spectra (UV-vis) were obtained for the re-dispersed samples (**Figure 5d**) which showed peaks at ~605 nm and ~660 nm attributed to the B and A excitonic transitions, respectively.[80] After functionalization, the B peak position remains the same while the A peak shifts toward longer wavelengths. While Eda et al.[80] have shown a redshift in both A and B peaks for thicker $MoS_2$ layers due to the change in band structure, we only see a redshift in the A peak, and we do not expect the functionalized $MoS_2$ to re-aggregate into thicker flakes because the 4-NBD groups prevent restacking. Instead, we postulate that the redshift can be attributed to electronic coupling of excitons to the conjugated aromatic groups attached to the $MoS_2$ surfaces. A similar exciton redshift has been observed for CdSe quantum dots coated in 4-mercaptobenzoic acid, which has a similar aromatic ring.[81] The increase in absorbance for functionalized $MoS_2$ below ~575 nm is attributed to absorbance of the attached organic groups.

Further characterization of the bulk $MoS_2$ dispersions with and without diazonium functionalization was conducted using thermogravimetric analysis (TGA) (see Methods for details). The TGA curves and first derivative curves are shown in **Figure 5e** and **Figure 5f**, respectively. The functionalized and unfunctionalized $MoS_2$ both show a small mass loss below 100°C, which may be due to residual adsorbed water and other small molecules. There is also some mass loss below 200°C, which is more prominent for the functionalized $MoS_2$, that may be due to removal of van der Waals bonded molecules, similar to the observation of Knirsch et al.[31] for the functionalization of lithiated phase-transformed $MoS_2$. The diazonium salt by itself has a significant mass loss by about 160°C (see Supporting Information, **Figure S18**), likely corresponding to a combination of sublimation and decomposition.

A sharp and prominent mass loss peak, corresponding to a mass loss of about 8%, occurs between 200°C to 300°C, with the maximum loss rate occurring at about 225°C. This sharp mass loss peak suggests the breaking of covalent bonds for species attached to the $MoS_2$ surface and the loss of nitrophenyl groups from the surface. There is also continued mass loss above 300°C, but it occurs at a faster rate, and with an additional peak at ~650°C and accelerating further above 800°C, which can be attributed to lattice degradation at higher temperatures. Using the total mass loss of the functionalized sample between 100°C and 450°C, which is about 15.7%, and using the molar masses of nitrobenzene and $MoS_2$, we estimate a surface coverage of about 24% assuming both sides of a monolayer $MoS_2$ flake in solution are available, or approximately 12% coverage on each side. This value is in direct agreement with the optimal concentration of ~8-12% predicted from DFT. This value is also comparable to the values estimated by Knirsch et al.[31] for the functionalization of lithium-treated phase-transformed $MoS_2$, which also showed the main mass loss due to covalently attached groups occurring between about 200 to 450°C. Thus, we have achieved a similar degree of functionalization without the use of lithium treatment using the defect-mediated cooperative mechanism. This temperature-dependent behaviour is also consistent with the removal of functional groups from mechanically exfoliated functionalized $MoS_2$ upon thermal annealing, as shown in the AFM images of Supporting Information, **Figure S19**, which shows that after annealing at 500°C the bound groups are largely removed.

*Protein attachment*

MoS$_2$ is an attractive material for biological applications because it is biocompatible,[82] and it provides a large surface area for functionalization and for interaction with biomolecules, and its semiconducting properties enable MoS$_2$ to be used for biomolecule sensing via changes in charge transport or fluorescence. The large area and 2D nature of MoS$_2$ also provide the potential for patterned arrays of devices for multiplexed detection schemes. Thus, examples of MoS$_2$ and TMDCs used in biological applications[26-28] include bioimaging,[46] drug delivery and treatment,[46,83] and biosensing.[84] Previous studies have highlighted the excellent properties of MoS$_2$ for biosensing by incorporating it into field-effect transistors (FETs) that provided substantial sensitivity improvements over graphene-based devices[85] and, using gate dielectrics functionalized with protein antibodies, detection limits down to 60 fM concentrations.[86] However, direct covalent attachment of proteins to the MoS$_2$ has the potential to dramatically improve device sensitivity by bringing protein-analyte interactions to the semiconductor surface.[87] The ability to covalently tether active proteins to the surface of semiconducting MoS$_2$ thus represents a critical step to realizing the full biosensing potential of MoS$_2$. Accordingly, we developed a diazonium-based chemistry to tether fluorescent proteins[59,60] to the MoS$_2$ surface. We first grafted carboxylic acid groups to the MoS$_2$ surface by functionalizing it with 4-carboxybenzenediazonium (4-CBD) tetrafluorobate (the UV-vis spectrum of bulk MoS$_2$ functionalized by 4-CBD is shown in the Supporting Information, **Figure S21**). The 4-CBD functionalized MoS$_2$ was then subsequently reacted to allow tethering of poly-histidine (His)-tagged green fluorescent protein (GFP) and the red fluorescent protein, mCherry (see Methods for protein synthesis and attachment details). The chemical attachment is schematically illustrated in **Figure 6a**. We used mechanically exfoliated MoS$_2$, as shown in the optical microscope image of **Figure 6b**. AFM imaging of the pristine MoS$_2$ shows smooth atomic steps (**Figure 6c**). After protein attachment, AFM imaging (**Figure 6d**) shows a uniform increase in thickness (**Figure 6e**) that we interpret as a layer of proteins attached via the Ni-chelating linkage shown in the schematic, along with some pinholes consistent with those observed for 4-NBD attachment in **Figure 3**.

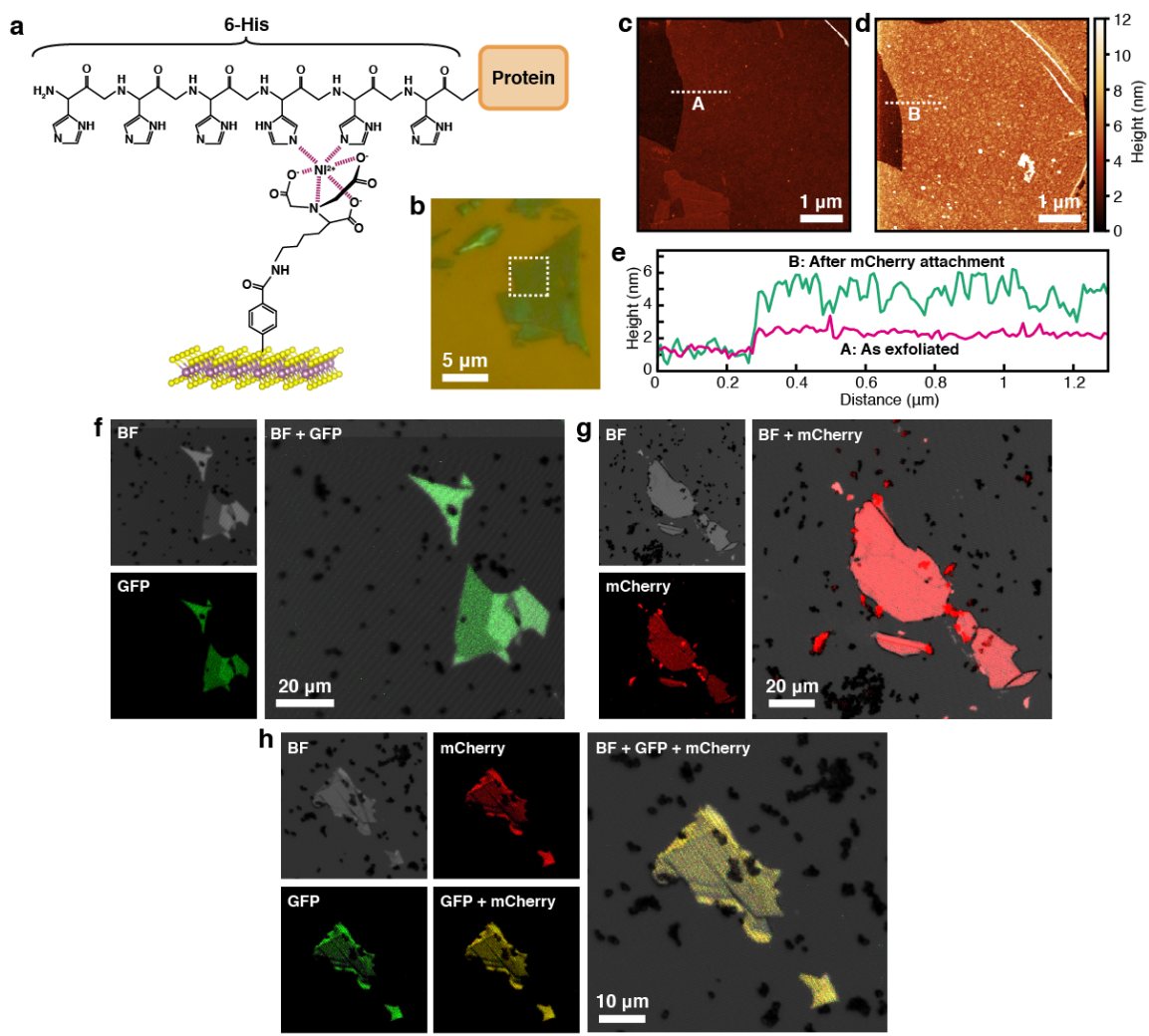

**Figure 6: Attachment of active proteins to MoS$_2$. (a)** Schematic of NTA-Ni-chelation attachment of poly-histidine (His) tagged protein, linked to MoS$_2$ surface via diazonium functionalization chemistry. The diazonium salt used here was 4-carboxybenzene diazonium (4-CBD) tetrafluoroborate. **(b)** Optical microscope image of mechanically exfoliated MoS$_2$ flakes featured in panels (c) and (d). **(c)** AFM image of pristine MoS$_2$ in the region indicated by the dashed square in panel (b). **(d)** AFM image in the same region as panel (c) after attachment of mCherry (red fluorescent protein) following initial 10 min functionalization with 4-CBD. **(e)** Height profiles along lines A and B in panels (c) and (d). **(f-h)** Bright field (BF), confocal fluorescence microscopy images in GFP (green) and mCherry (red) channels, and fluorescence images overlaid onto BF images, after protein attachment process. **(f)** GFP attachment. **(g)** mCherry attachment. **(h)** 1:1 mixture of GFP and mCherry attachment. In panels (f-h), fluorescence is only observed at the locations of the MoS$_2$ flakes where the initial diazonium functionalization has formed the anchors for the attachment chemistry as illustrated in the schematic.

To confirm attachment of active, viable proteins, we used confocal fluorescence microscopy to image the samples. **Figures 6f-g** show bright field optical images along with fluorescence images in the green and red channels for GFP and mCherry emission, respectively. These images indicate that the proteins have been successfully attached to the MoS$_2$ regions where the initial diazonium functionalization took place, and not in regions of the bare SiO$_2$/Si substrates. Furthermore, the strong fluorescence signals that we detect from the proteins demonstrate that our attachment chemistry is sufficiently gentle to avoid protein denaturation. **Figure 6h** shows the result of attaching a 1:1 mixture of GFP and mCherry. The overlay of both red and green channels on the bright field image, resulting in a combined yellow appearance, shows that both proteins have been uniformly localized to the $MoS_2$ flakes. Additional AFM and confocal images of more protein-functionalized samples are shown in the Supporting Information, **Figure S20**. We also note that in the samples shown in both **Figure 6** and **Figure S20** that the thicker regions of the $MoS_2$ flakes show brighter fluorescence, which we interpret to mean that a higher concentration of initial diazonium attachment sites results in a higher concentration of fluorescent proteins, consistent with the denser grafting of groups shown in AFM images in **Figures 3 and S13**, and DFT calculations in **Figure 2**. In a control experiment where the diazonium functionalization step is skipped but all other processing steps are kept the same, no fluorescence is detected. Thus the covalent 4-CBD attachment step is crucial, and any possible association of the $Ni^{2+}$ ions to the $MoS_2$ surface such as via coordination chemistry[34] is not sufficient to tether the proteins.

*Discussion of reaction mechanism and coverage levels*

The diazonium reaction mechanism for $MoS_2$ differs from the mechanism for functionalization of other materials. The process of covalent functionalization with aryl diazonium salts in general relies on the transfer of an electron from the substrate to the diazonium molecule to generate a reactive radical. For graphene and carbon nanotubes, the electron transfer rate depends on the overlap of the substrate's occupied density of states (DOS) and the molecule's unoccupied states.[60,88-90] However, in the case of $MoS_2$, our DFT calculations show that the completely perfect $MoS_2$ lattice is not reactive (see **Figure S22** in the Supporting Information for DOS calculation), but the presence of a single sulfur vacancy will make the surrounding area reactive for covalent functionalization, which in turn generates another region of increased reactivity so that that subsequent molecules can react adjacent to existing ones. Thus, a single vacancy defect can allow the entire surface of otherwise perfect $MoS_2$ to be covalently functionalized, with the reaction spreading in a chain-like configuration. This type of cooperative effect, where molecules strengthen their covalent interactions with another molecule or surface when other molecules bind, has been observed for many biopolymers such as proteins and nucleic acids,[91-93] but no reports on such mechanisms have been addressed so far on the functionalization of the basal plane of TMDCs by simple diazonium molecules. We have also shown that adding a small concentration of vacancies by Ar plasma treatment increases the spread and density of the reacted groups by providing more initial sites of reaction.

We also observed in experiments and DFT calculations differences in the reactivity of $MoS_2$ as a function of layer number. Although layer number dependent reactivities have been reported in graphene, the behavior that we observed with $MoS_2$ was opposite that of graphene. For graphene, monolayer samples are much more reactive than bilayer and multilayer graphene:[89,94] charged impurities in the substrate induce small, locally n-doped regions that provide the necessary charge for the electron transfer step, but are screened by the first layer of graphene in bilayer and multilayer samples.[60] In contrast, bilayer and multilayer $MoS_2$ appears to be more reactive than monolayer $MoS_2$, with a higher density of attached groups observed by AFM (**Figure 3**) and a brighter fluorescence after protein attachment (**Figure 6**). In the case of $MoS_2$, our DFT calculations show that the presence of surface defects induces an increased surface dipole moment that contributes to increasing reactivity by layer number. Other reports have shown that both mechanically exfoliated and CVD-grown material is typically n-type because of sulfur vacancies,[69,95] and the DOS of the multilayer is three times higher than that of the monolayer, also contributing to the increased reactivity of multilayer $MoS_2$.[11]

The coverage of covalently functionalized sites was calculated from the TGA mass loss in **Figure 5** to be ~12%, which aligns with our DFT calculations for an optimal coverage of ~8-12%. For graphene, it has been estimated via first principles calculations that the maximum packing density of aryl groups on the surface is ~11% due to steric hindrance.[96] This value is very similar to our estimate based on TGA data, as well as that of Knirsch et al.[31] for functionalized 1T-$MoS_2$. In the AFM images of molecular coverage (**Figure 2**), each protrusion may represent more than one aryl group, and may be broadened due to the curvature of the AFM tip. Thus, the dense appearance of molecular groups at longer reaction times in AFM images is not directly related to the quantitative estimate of the density of C–S bonds. There may also be the formation of oligomers,[62] where the aryl radical attaches to an existing bound group rather than a bare $MoS_2$ site, which was discussed above in relation to the XPS data.

## Conclusion

In summary, we have demonstrated the direct covalent functionalization of the basal planes of pristine semiconducting 2H-$MoS_2$ using aryl diazonium salts without the use of any pre-treatments. The functionalization occurs in mild conditions, is very versatile, and can be applied effectively to $MoS_2$ obtained from different methods including mechanically exfoliated flakes, CVD-grown films, and solution phase dispersions that can be prepared in bulk quantities. Detailed DFT calculations reveal the reaction mechanism, dependence on layer number, and the origins of the observed chain-like growth mode, which relies on a novel cooperative surface reaction mechanism. The formation of covalent bonds is confirmed by XPS and FTIR, and surface coverage of bound groups is estimated by AFM and TGA. The resulting functionalized $MoS_2$ remains in its semiconducting 2H form, exhibiting a strong PL signal that reveals evidence of increased n-doping with higher levels of functionalization, providing a straightforward chemical route to engineering the Fermi level of the two-dimensional semiconductor. Furthermore, this aryl diazonium covalent chemistry is stable and robust, and we have extended it toward the tethering of active proteins, an enabling step for future biosensing applications. Our studies establish important fundamental findings in the chemistry of TMDCs, demonstrating that lithium-based phase conversions or aggressive chemistries with toxic metal chlorides are in fact not necessary for robust covalent functionalization, and that the maximum concentration of functionalized sites is not limited by the initial concentration of defect sites. These results also have important implications for using chemical functionalization to engineer the electronic and chemical properties of TMDCs. They provide a versatile platform for covalent chemical modification of $MoS_2$ using a diverse range diazonium salts and facilitate new methods of interfacing semiconducting $MoS_2$ with biomolecules for biological applications. There is tremendous potential for the broader chemical functionalization of TMDCs of other compositions based on these results.

**Acknowledgements**

We gratefully acknowledge the use of facilities at the LeRoy Eyring Center for Solid State Science at Arizona State University and Dr. Tim Karcher for assistance with XPS and TGA measurements; the Keck Bioimaging Lab at ASU and Dr. Page Baluch for assistance with confocal imaging; Prof. Hao Yan for use of the AFM and Raman systems, and Dr. Shuoxing Jiang for assistance with AFM measurements, and Dr. Su Lin for assistance with Raman measurements. X.S.C., A.Y., Q.H.W., and A.A.G. acknowledge support from ASU startup funds and NSF grant DMR-1610153. A.A.G. acknowledges an Alfred P. Sloan Research Fellowship (FG-2017-9108) and an Arizona Biomedical Research Commission New Investigator Award (ADHS16-162400). E.J.G.S. acknowledges the use of computational resources from the UK National Supercomputing Service, ARCHER, for which access was obtained via the UKCP consortium and funded by EPSRC grant EP/K013564/1; the Extreme Science and Engineering Discovery Environment (XSEDE), supported by NSF grants TG-DMR120049 and TG-DMR150017; and the Queen's Fellow Award through the startup grant number M8407MPH and the Enabling Fund (QUB, A5047TSL).

**Author Contributions**

X.S.C. performed the chemical functionalization experiments, time dependent Raman and PL measurements, AFM measurements, XPS analysis, STM measurements, and Ar plasma experiments. A.Y. performed the 4-carboxybenzenediazonium tetrafluoroborate synthesis, protein attachment experiments, FTIR and UV-vis measurements. D.O.L. performed the CVD growth, and additional chemical functionalization, Raman, PL, and AFM measurements. A.A.T. performed the protein synthesis and purification. A.Y. and A.D. performed the solution phase dispersion and functionalization experiments. A.Y. and D.M. performed the confocal microscopy measurements. A.D. performed NMR characterization of synthesized diazonium salts and prepared samples for TGA. E.J.G.S. performed the DFT calculations and analysis. X.S.C., E.J.G.S., and Q.H.W. analyzed the data and wrote the manuscript. Q.H.W., E.J.G.S. and A.A.G. supervised the project. All authors discussed the results and revised the manuscript.

**Competing Financial Interests statement**

The authors declare no competing financial interests.

**Supporting Information**

Additional Raman and PL spectra and spatial maps, more details of DFT results, additional AFM and STM images, wide scan XPS spectra, additional UV-vis spectra and TGA curves, nitrobenzene control experiments, characterization of CVD-grown samples, morphology of solution phase derived films.

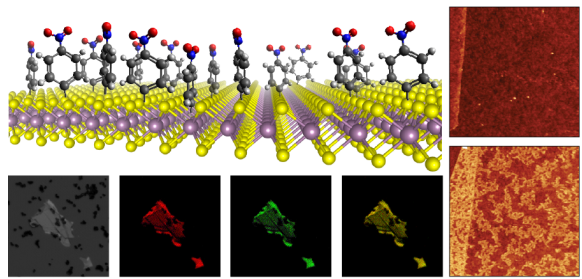

**Table of contents figure**